# Dynamic Video Recognition from Water Chip

***Kangchen Xiong[1,#], Minhui Yang[1,#], Xiaokai Bai[1,#], Tengfei Wu[2], Wanyu Zhang[2], Haoxin Wu[1], Shisheng Lin[1,3,*]***

[1]College of Information Science and Electronic Engineering, Zhejiang University, Hangzhou, 310027, P. R. China.

[2]College of Integrated Circuits, Zhejiang University, Hangzhou, 310027, P. R. China.

[3]State Key Laboratory of Extreme Photonics and Instrumentation, Zhejiang University, Hangzhou, 310027, P. R. China.

## Abstract

Dynamic video recognition requires efficient extraction of spatiotemporal information from evolving scenes. Conventional CMOS image sensors read out pixels independently, requiring intensive downstream calculations to extract correlations across pixels and frames. Here we exploit the intrinsic physics of a hydrogen bonded water molecule field to couple photoactive pixels and preprocess optical inputs through physical integration before readout. A Green's function formulation of the polarization transfer function (PTF) describes how light-induced responses from different regions are spatially weighted and superposed, with a nonlocal response kernel spanning the photoactive area. In a graphene-hydrogel-silicon chip, a patterned graphene photoactive array produces complementary current-time signals collected through eight peripheral readouts at zero external bias. This readout parallels neural population decoding in brain computer interfaces, where distinct signal patterns across electrodes distinguish different movements. A multiscale temporal neural network decodes these signals to recognize moving light sources, evolving contours and internal patterns, including low contrast actions, ambient light gestures and human actions in dense fog. By harnessing the natural physical integration of the water molecule field, this approach offers a route towards low power, noise tolerant recognition of dynamic scenes.

## Main Text

Dynamic video recognition underpins autonomous robotics, human-machine interaction and edge computing. Despite advances from conventional motion analysis[1] to three-dimensional convolutional networks[2,3] and video transformers[4], most systems still acquire dense image sequences before extracting task-relevant features. This acquisition-first paradigm incurs substantial data movement and memory overhead, contributing to the energy and latency costs of visual processing[5,6]. Network compression and low precision inference can reduce downstream arithmetic, but cannot eliminate the cost of pixel-wise acquisition and data transfer. Physically coupling photosensitive elements to preprocess optical information before readout offers a more direct route to reducing the computational, bandwidth and energy demand of visual recognition.

Event cameras reduce frame redundancy through asynchronous intensity-change detection but retain pixel-addressed outputs[7]. Two-dimensional material image sensors and in-sensor processors, bring classification and motion analysis closer to the detector, although many still rely on dense arrays or programmable pixel states[8-19]. In parallel, sparse boundary electrodes have enabled distributed thermal, tactile sensing[20,21], and optical spot localization[22]. Recent non pixelated optoelectronic sensors further demonstrate that peripheral electrical measurements can support spatiotemporal perception[23,24]. Motor brain computer interfaces offer a related example of sparse electrical readout. In an intracortical interface for handwriting, electrodes record local voltage fluctuations as a participant attempts to write prompted characters[25,26]. A decoder learns the correspondence between patterns of neural activity across channels and the prompted characters, then uses that mapping to infer what the participant attempts to write in a new trial. An open question is whether a shared medium can physically integrate photoresponses across a sensing area so that a few peripheral electrodes can capture scene dynamics at zero bias.

Our previous work demonstrated self-driven photodetection in graphene-polar liquid-semiconductor heterostructures, where photoinduced interfacial polarization generates transient currents at zero bias[27–30]. The hydrogen bonded water molecule field provides an interconnected polar environment for coupling spatially separated photoresponses. We subsequently established the polarization transfer function (PTF), revealing lateral signal transfer, exponential spatial attenuation and superposition under stationary illumination[31]. The PTF admits a nonlocal Green's-function description in which distant optical perturbations contribute to the same electrical readout. Whether this field-wide weighted integration can provide an invertible encoding of continuously evolving spatiotemporal optical fields remains an open question.

Here we extend PTF-mediated coupling to dynamic vision using graphene-hydrogel-silicon arrays with eight light shielded peripheral readouts. We test whether polarization transfer through a shared polar medium can integrate spatially distributed photoresponses into a small set of current traces at zero external bias. The water chip shares a readout principle with motor brain computer interfaces. During attempted handwriting, electrodes sample neural activity at different sites in the motor cortex. In the chip, each peripheral electrode records a different spatially weighted sum of photoresponses across the illuminated window. In both systems, the relative amplitudes and timing of signals across electrodes help the decoder identify the character or action in a new trial. Our CNN-BiLSTM classifier distinguishes motion, evolving contours and internal patterns, including low-contrast actions, ambient-light gestures and human actions in dense fog. Operational stability and tolerance to substantial random RGB noise further establish the operating scope of the device. By exploiting the PTF of the shared water molecule field for physical preprocessing before readout, this architecture offers a route towards low power, noise tolerant recognition of dynamic scenes.

**Device architecture and sparse peripheral readout**

Fig. 1a illustrates the device architecture and electrical acquisition system. An 8 × 8 $cm^2$ silicon substrate supports a coextensive hydrogel layer and a central graphene array spanning 5 × 5 $cm^2$. The shared hydrogel enables lateral transfer of photoinduced interfacial polarization, physically coupling spatially separated pixels whose contributions to the peripheral currents are described by a Green's function based weighted superposition. Eight 1.5 × 1.5 $cm^2$ signal reading units are positioned at the four corners and four edge midpoints. An opaque encapsulation layer surrounds the photosensitive region, preventing direct illumination and limiting hydrogel dehydration. The eight readout units are sequentially routed through an 8:1 multiplexer to a Keithley 6514 electrometer.

Fig. 1b summarizes the motion recognition workflow. Nine projected spot trajectories induce local changes in interfacial polarization that couple laterally through the hydrogel, generating current sequences with distinct amplitudes and temporal structures. After preprocessing, a multiscale CNN-BiLSTM classifies the signals, as summarized by the nine class confusion matrix. The PTF spatially weights and naturally superposes optical inputs within the material, mapping a dynamic optical field onto a small number of complementary peripheral temporal signals and thereby performing spatial information compression at the sensing front end while reducing subsequent readout and computational complexity.

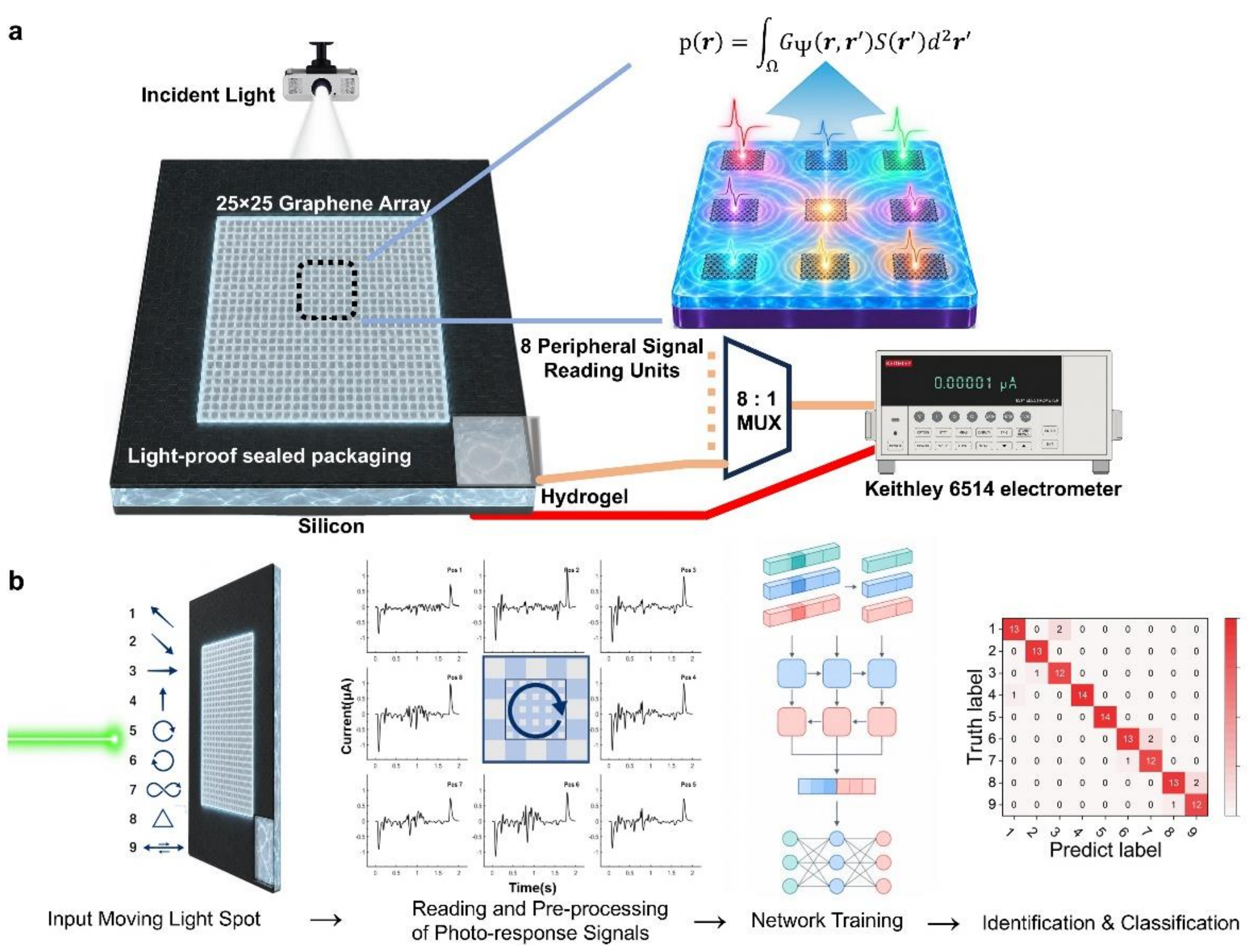

**Fig. 1 | Self-driven dynamic vision device with interactive PTF-mediated sparse peripheral readout.**

**a**, Schematic of the device and acquisition system.

**b**, Sensing and classification of dynamic optical inputs.

## Photoinduced interfacial polarization and distance dependent transfer

Each peripheral readout comprises PET-supported graphene with an Ag contact, a hydrogel layer and n-type silicon with Au back electrode (Fig. 2a). Photoinduced interfacial polarization couples laterally through the shared hydrogel, generating transient currents at the peripheral readouts.

To characterize this lateral transfer, localized illumination was applied at increasing distances from the readout (Fig. 2b). Light onset and removal generated opposite polarity transients, with exponentially decreasing amplitude and increasing temporal broadening at increasing distances (Fig. 2c).

$$A(d) = A_0 \exp\left(-\frac{d}{\lambda}\right) \qquad (1)$$

where $d$ is the lateral source-to-readout distance, $A_0$ is the local amplitude, and $\lambda$ is the characteristic lateral polarization transfer length. This centimeter scale PTF allows each peripheral unit to receive polarization fields from macroscopic scale. Exponential lateral attenuation and natural superposition of the PTF were established in the preceding polar liquid platform[31].

Figures 2d-f illustrate the proposed interfacial polarization mechanism. Before contact, graphene and n-type silicon have different Fermi levels (Fig. 2d). Upon equilibration, interfacial charge redistribution induces band bending and orientational polarization of polar molecules and interfacial dipoles in the hydrogel (Fig. 2e). Optical absorption in silicon promotes electrons from the valence band to the conduction band, modifies the interfacial carrier density and local electric field, and enhances the polarization of the polar medium (Fig. 2f). After light removal, the photocarrier population decreases and the polarization relaxes towards its dark state configuration, producing a displacement current of the opposite polarity. This interpretation follows the polarization enhancement and relaxation mechanism established for the preceding graphene-polar liquid-semiconductor detector[27–30]. The electric displacement $\mathbf{D}$ and displacement current density $\mathbf{j}_{\mathrm{D}}$ satisfy

$$\mathbf{D} = \varepsilon_0 \mathbf{E} + \mathbf{P}, \qquad \mathbf{j}_{\mathrm{D}} = \frac{\partial \mathbf{D}}{\partial t} = \varepsilon_0 \frac{\partial \mathbf{E}}{\partial t} + \frac{\partial \mathbf{P}}{\partial t} \qquad (2)$$

where $\mathbf{E}$ is the local electric field, $\mathbf{P}$ is the polarization density and $\varepsilon_0$ is the vacuum permittivity. Under constant illumination, the polarization approaches a new quasi-equilibrium state and the displacement current returns towards baseline.

To extend the PTF from stationary optical switching to continuously evolving inputs, we describe the underlying physics of the polar medium in terms of a hydrogen bonded water molecule field driven by local photoinduced interfacial polarization. Let $\Omega$ denote the photosensitive domain, $\mathbf{r} = (x, y)$ an in-plane position and $I(\mathbf{r}, t)$ the local optical intensity at time $t$, encompassing the input's geometry, brightness and temporal evolution. In the unsaturated linear response regime, the effective photoinduced polarization source is

$$S(\mathbf{r}, t) = \chi(\mathbf{r}) M_g(\mathbf{r}) I(\mathbf{r}, t). \qquad (3)$$

Here $M_g(\mathbf{r})$ is the dimensionless structural weighting of the graphene layer, and $\chi(\mathbf{r})$ is the local normal interfacial polarization change per unit optical intensity, which accounts for variations in absorption, band alignment, interfacial states and polarization efficiency across the photoactive elements.

For peripheral readout $i$, centred at $\mathbf{r}_i$ with $i = 1, \ldots, 8$, the source-to-readout distance is

$$d_i(\mathbf{r}) = \|\mathbf{r} - \mathbf{r}_i\|_2 \qquad (4)$$

The spatial attenuation and distance dependent temporal broadening, quantified by the transient width $W(d)$, together define $G_\Psi(\mathbf{r}_i, \mathbf{r}; \tau)$ as the causal current response kernel to a unit localized impulse of the polarization rate source. Here $\tau = t - t'$ is the elapsed time since excitation at $t'$, and $\Psi$ labels the water molecule polarization field. We parameterize the Green's function using the measured PTF as

$$\begin{aligned} G_{\Psi}(\boldsymbol{r}_i, \boldsymbol{r}; \tau) \; &\simeq \; \exp\left[-\frac{d_i(\boldsymbol{r})}{\lambda}\right] h(\tau; W[d_i(\boldsymbol{r})]) \\ W(d) \; &= \; W_0 \; + \; \beta d^{\zeta} \end{aligned} \tag{5}$$

The characteristic lateral transfer length is $\lambda$; $W_0$ is the near field response width, $\beta > 0$ the broadening coefficient and $\zeta > 0$ a fitted exponent. The function $h$ vanishes for $\tau < 0$ and describes the transient waveform produced by switching a stationary localized optical input. Its temporal profile reflects the device band structure, polar layer composition and thickness, polarization relaxation and effective interfacial capacitance.

The current $P_i(t)$ is obtained by integrating the local polarization changes over the photosensitive domain and their causal history:

$$P_i(t) \; = \; -\int_0^t \int_{\Omega} G_{\Psi}(\boldsymbol{r}_i, \boldsymbol{r}; t - t') \; \frac{\partial S(\boldsymbol{r}, t')}{\partial t'} \; \mathrm{d}^2\boldsymbol{r} \; \mathrm{d}t' \tag{6}$$

Here $\mathrm{d}^2\mathbf{r}$ is the in-plane area element, and the negative sign specifies the measurement polarity. Equation (6) extends the spatial Green's function integral to continuous dynamic optical field: local illumination changes drive polarization in the water molecule field, and their spatially weighted responses are physically integrated at each electrode. Distinct electrode positions therefore provide complementary temporal projections before downstream digital processing.

Fig. 2g illustrates this encoding using a spot translating from left to right. Spot appearance near the left readout unit produces a pronounced negative transient, whereas the distant right unit records a weaker and broadened response. During translation, polarization enhancement at the leading edge and relaxation at the trailing edge coexist and partially cancel, generating lower amplitude fluctuations. Spot disappearance near the right unit subsequently produces a large positive transient. The left and right traces therefore constitute complementary temporal projections of the same moving input. Even optical perturbations approximately 7 cm from a readout electrode produce detectable responses, revealing the long-range reach of polarization transfer. This device-scale coupling gives each peripheral current a spatially weighted projection of the entire photoactive field without individual pixel acquisition.

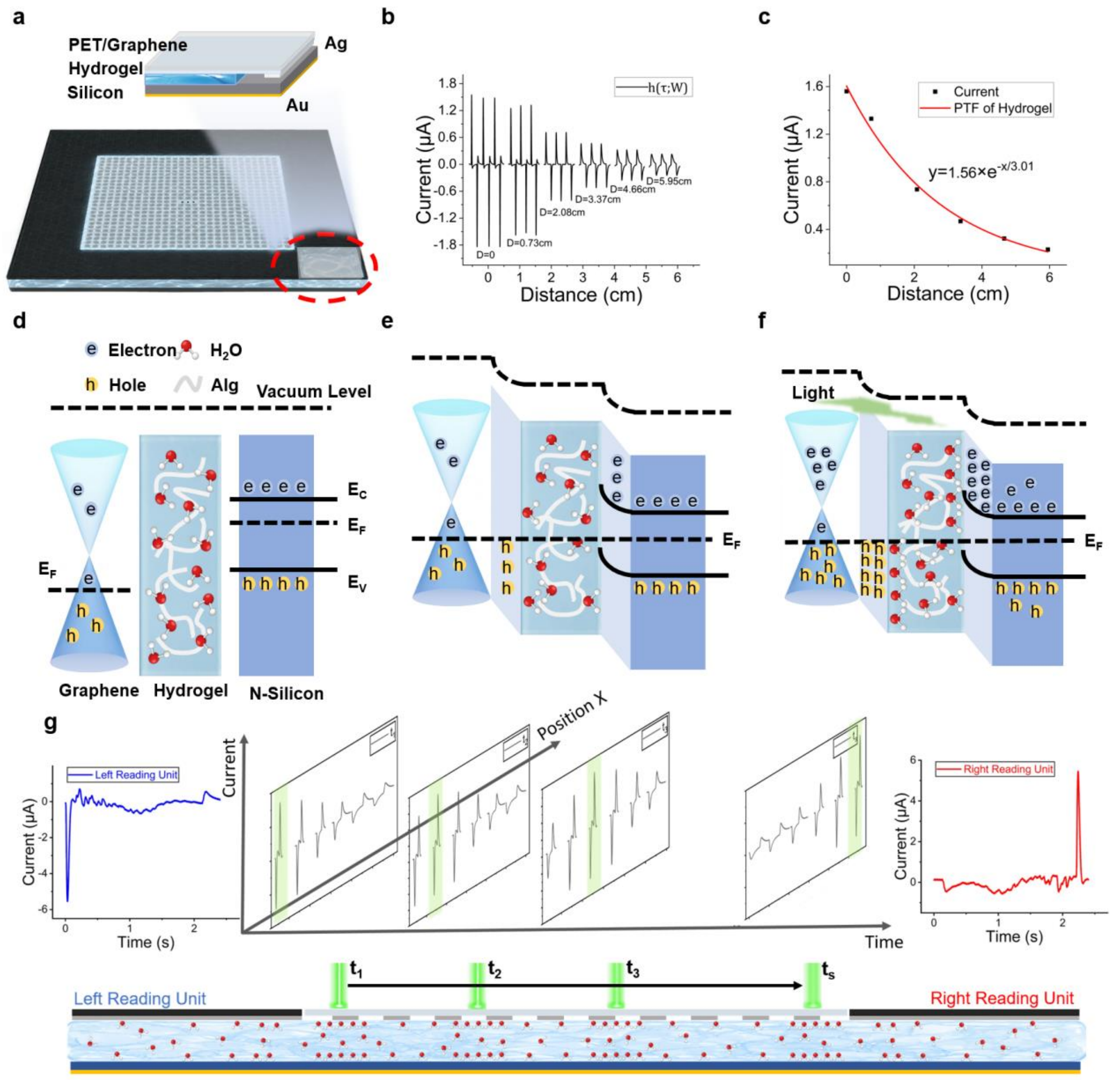

**Fig. 2 | Photoinduced interfacial polarization, distance dependent polarization transfer and peripheral responses to a moving optical input.**

**a,** Structure of a peripheral signal reading unit and its position within the complete device.

**b,** Bipolar current transients generated by switching a stationary localized optical input at source-to-readout distances $\mathrm{d} = 0, 0.73, 2.08, 3.37, 4.66\,, 5.95$ cm.

**c,** Distance dependence of the extracted characteristic current amplitude.

**d-f**, Schematic band diagrams of the graphene-hydrogel-Si heterostructure: d, before contact; e, after contact and equilibration, when the Fermi-level mismatch drives interfacial charge redistribution and orientational polarization; and f, under illumination, when photogenerated carriers in silicon enhance the interfacial polarization.

**g,** Spatiotemporal response to a light spot translating from left to right. The lower schematic shows the spot positions at $t_1 < t_2 < t_3 < t_s$, and the central sequence illustrates the local and transferred transient contributions generated at selected positions. Blue and red traces show the current–time responses of the left and right reading units, respectively.

## Multiscale temporal decoding and the contribution of peripheral channels

The eight peripheral traces were decoded using a multiscale CNN-BiLSTM network (Fig. 3a). Two smoothing scales captured fine and coarse temporal variations, followed by convolutional feature extraction and BiLSTM-based temporal integration.

Figures 3b, c compare selected configurations of one to eight peripheral channels.

Increasing channel number generally reduces loss and improves convergence stability and accuracy, with diminishing gains at higher channel counts $N_{ch}$. These trends suggest that spatially separated PTF signals carry complementary but partially overlapping information.

**Temperature, temporal stability, motion speed and optical corruption**

The PTF-derived temporal signatures remained largely preserved under moderate temperature variations and device ageing (Fig. 3d, e). At 25-45 °C, the dominant peaks, valleys and temporal envelopes showed limited changes, with a fixed 25 °C-trained classifier maintaining 86.67-88.89% accuracy. At 55 °C, waveform distortion reduced accuracy to 73.33%, while pronounced transient deformation and baseline shifts at 65 °C lowered it to 47.78%. Despite minor amplitude and baseline drift, recognition remained at 94.44-97.78% over device ages of approximately 25-40 days using the initially trained model, which shows the long-term operational stability of the device.

Motion speed reshaped the separation, frequency and overlap of the response features (Fig. 3f). Nevertheless, independently trained models achieved 100% accuracy across six speed factors from 1× to 20×, indicating that class-specific features remained distinguishable over these timescales.

Fig. 3g-i assess the robustness of the device to progressively intensified RGB noise. Increasing noise produced a systematic loss of image quality, with the SNR decreasing from 10.82 to -5.04 dB and the SSIM from 0.977 to 0.340 across levels L1-L5 (Fig. 3g, h). Although the initial and terminal extrema of the current transients were progressively attenuated, the intermediate temporal features of the current responses remained well resolved (Fig. 3i). The recognition accuracies decrease from 100.00% to 84.44% over L1-L4 and remain 71.11% even at L5. These results demonstrate that PTF-mediated recognition remains robust under substantial RGB noise, provided that the discriminative temporal structure is preserved.

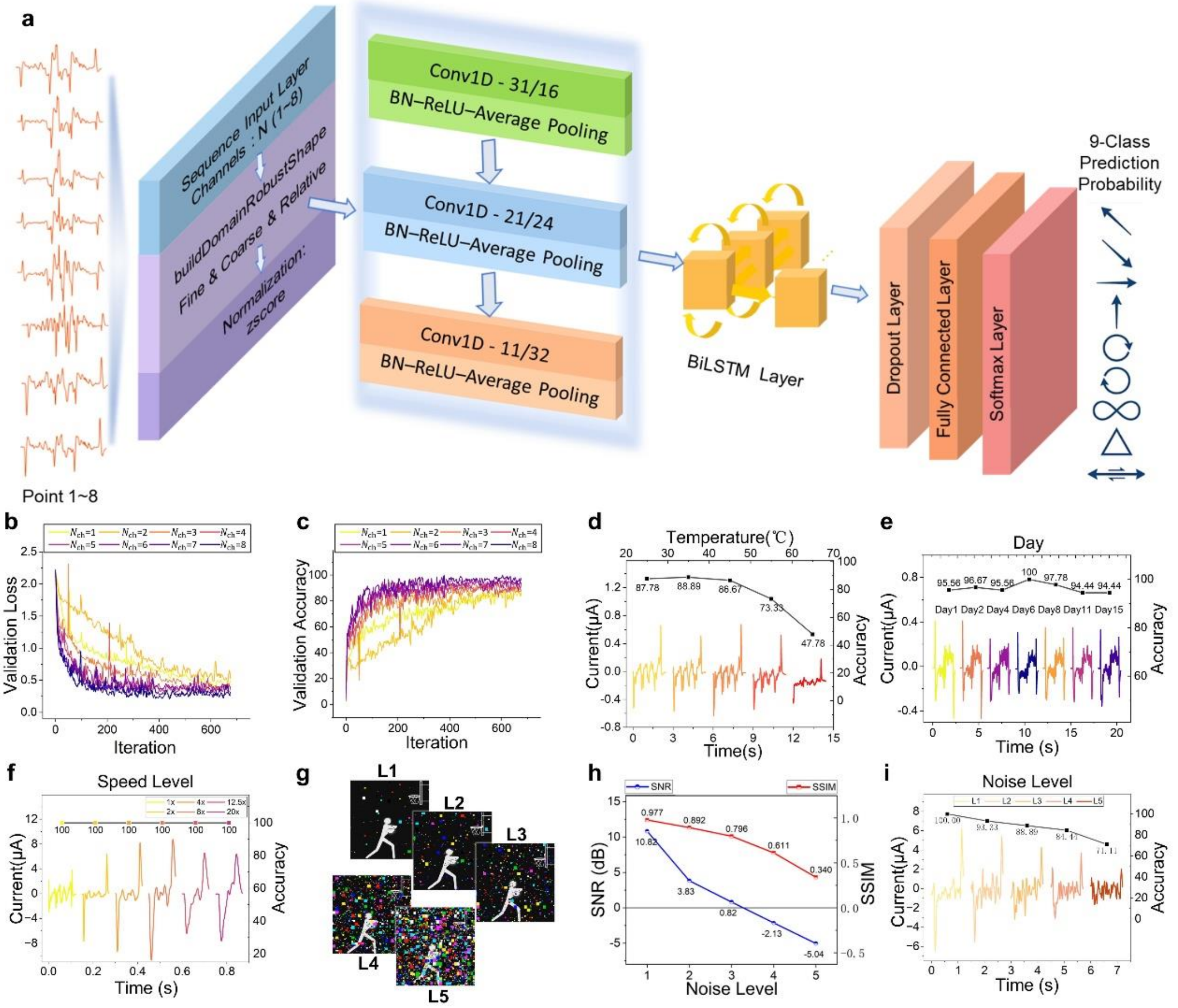

**Fig. 3 | Multiscale temporal decoding and recognition under different operating conditions.**

**a,** Multiscale shape CNN-BiLSTM architecture.

**b-c,** Validation loss and validation accuracy, respectively, as functions of iteration for models using different numbers of peripheral channels.

**d,** Representative responses and classification accuracy at different device-surface temperatures using a fixed model trained at 25 °C.

**e,** Long-term stability from device age 25 days to 40 days, evaluated using a fixed model trained on the initial dataset.

**f,** Representative responses and classification accuracy at six motion speeds; an independent model is trained at each speed.

**g,** Representative input images under progressively increasing RGB noise levels (L1-L5).

**h,** Image quality metrics (SNR and SSIM) of different RGB noise levels.

**i,** Current responses under different RGB noise levels and the corresponding recognition accuracy.

## Recognition of complex dynamic scenes and X-ray patterns

To determine whether the peripheral signals retain local spatial information beyond the global position and trajectory of objects, we examined tasks spanning dynamic contours, internal patterns, low contrast scenes and passive ambient light gestures (Fig. 4a-h).

We first examined nine animated sporting silhouettes, in which the centroid, limb configuration, contour and illuminated area evolved over time (Fig. 4a, b). All channel configurations learned the action classes rapidly, with multichannel models generally converging more smoothly towards high validation accuracy, indicating that the signals support recognition of articulated, changing silhouette shapes. Another test used nine national flags with matched global waving kinematics but distinct color regions, boundaries and graphic pattern (Fig. 4c, d). Their successful discrimination shows that the signals encode not only global motion direction and centroid trajectory but also time varying internal structure.

Fig. 4e-f extend the task to camouflage-like human actions in which the foreground and forest background have similar color and intensity. The resulting weak local modulation limited the information available to individual readout points: one- and two-channel models improved only gradually, whereas models using five to eight channels reached substantially higher and more stable validation accuracy. The pronounced $N_{ch}$ dependence indicates that low contrast scenes require complementary PTF projections from multiple spatial directions: changes that approach the noise floor in one channel may remain detectable in another channel with a different source-to-readout geometry.

We performed another test to evaluate near field physical object motion by moving hands near the device, which occluded and redistributed ambient illumination across the photosensitive area (Fig. 4g, h). The resulting peripheral responses enabled the five gestures to be classified with high validation accuracy using one to four readout channels. This result demonstrates near field human machine interaction through passive ambient light modulation, without an additional illumination source.

We further evaluated six actions from the KTH Action Recognition Dataset[32] ( Fig. 4i, j) and nine actions from FogAct[33] (Fig. 4k, l). For KTH, validation accuracy increased from approximately 60-65% with one or two channels to 85-95% with six to eight channels. FogAct yielded lower accuracies overall, reaching approximately 90% with seven or eight channels, compared with approximately 40% with one or two channels. Both datasets showed an overall, although non-monotonic, benefit from multichannel readout, extending action recognition to recorded scenes with varied performers and environmental interference.

Together, the results show that PTF-mediated peripheral readout preserves task-relevant contour, pattern and contrast information beyond global trajectory. Channel diversity supplies complementary spatial projections, improving discrimination when individual responses are weak or ambiguous. Recognition across varied performers, backgrounds and visibility conditions further suggests the potential of device for action monitoring in complex real-world environments.

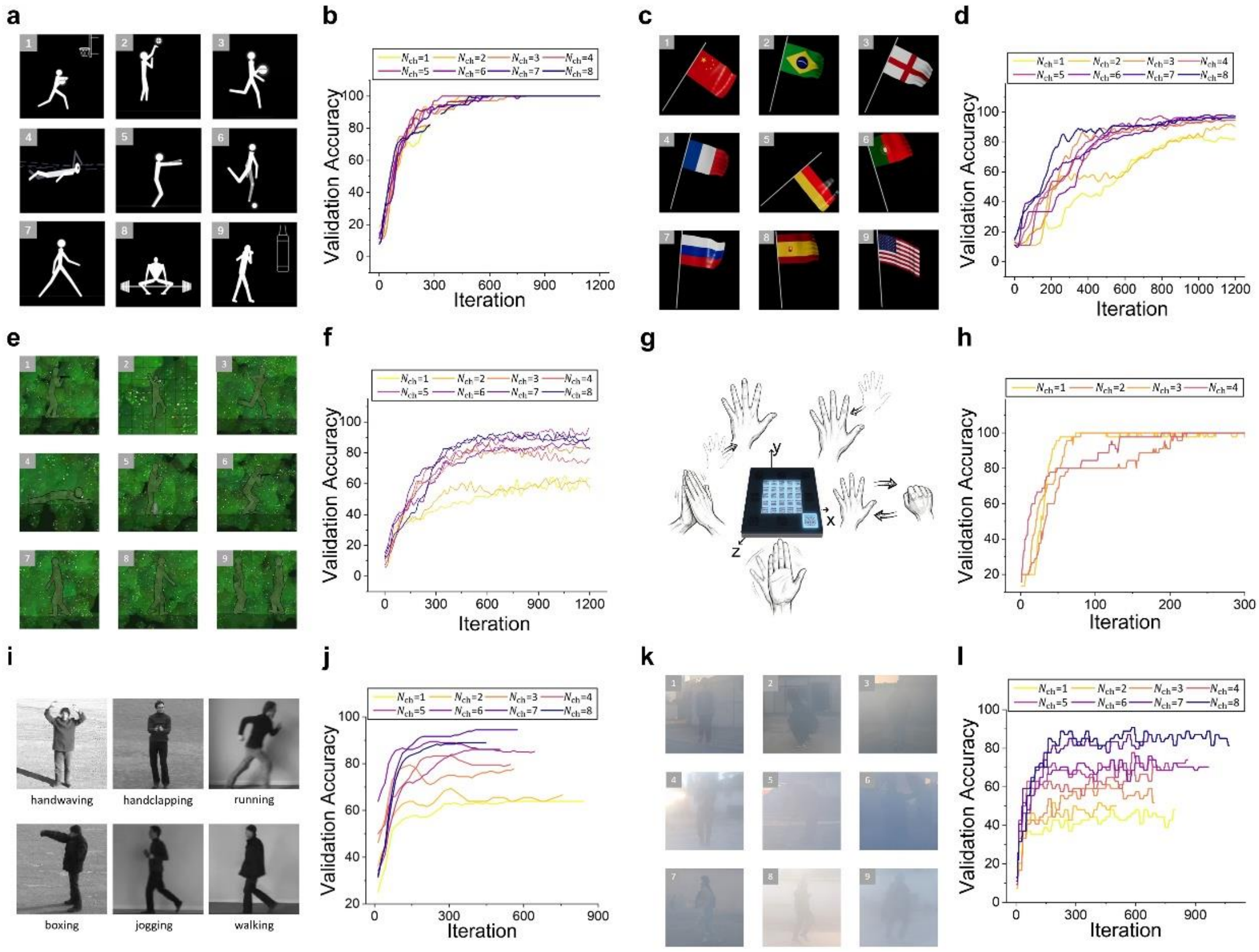

**Fig. 4 | Recognition of complex dynamic scenes, ambient-light gestures and direct X-ray patterns using sparse peripheral readout.**

**a,** Representative frames from nine animated sporting actions silhouettes.

**b,** Validation accuracy as a function of iteration for sporting actions classification using one to eight peripheral channels.

**c,** Representative frames from nine national flag animations. The classes share the same or matched global waving kinematics.

**d,** Validation accuracy for flag classification using one to eight peripheral channels.

**e,** Nine camouflage-like human actions classes with low foreground-background color and intensity contrast.

**f,** Validation accuracy for camouflage actions recognition.

**g,** Schematic of ambient light gesture recognition. The five gestures are: reaching in, reaching out, opening and closing the palm, waving, and clapping.

**h,** Validation accuracy for four gesture classes using one to five reading channels.

**i,** Representative frames of six actions from the KTH Action Recognition Dataset: hand waving, hand clapping, running, boxing, jogging and walking.

**j,** Validation accuracy versus training iteration for KTH action recognition using one to eight peripheral readout channels.

**k,** Representative frames of nine actions from the FogAct dataset under fog-degraded visibility: Jumping, Squatting, Moving Away, Approaching, Sweeping, Opening Umbrella, Running Left, Running Right, Standing up.

**l,** Validation accuracy versus training iteration for FogAct action recognition using one to eight peripheral readout channels.

## Toward material-level machine vision

**Physical encoding before Calculation.**

PTF provides a route to process visual information before pixel-wise digitization. Rather than treating each photosensitive element as an independent output, the sensing medium spatially weights and superposes distributed optical perturbations, converting a high-dimensional dynamic field into a small set of temporal electrical representations. The photosensitive layer therefore acts not only as a detector, but also as a physical encoder that performs part of the feature formation intrinsically.

**Co-design across materials and computation.**

In this framework, polarization-transfer length, temporal relaxation, device geometry and readout position define the encoding kernels of the sensor. These parameters can therefore be engineered jointly with the downstream temporal model. Regulating the polar medium, graphene patterning and peripheral architecture may enable task specific physical projections, shifting dimensionality reduction from purely digital processing towards the sensing material itself.

**From device physics to edge intelligence.**

The next step is to translate this principle into integrated hardware through simultaneous low noise acquisition, compact analogue front ends and lightweight temporal inference. System level energy, latency and information throughput should be evaluated within a unified hardware boundary. By modifying the photoactive layer or energy conversion medium, the same concept may also extend beyond visible light motion sensing. More broadly, material-mediated spatial coupling offers a general strategy for task-oriented perception where sensing and information processing are no longer strictly separated.

## Conclusions

We establish polarization transfer as a material-level mechanism for encoding dynamic optical fields. A structured graphene-hydrogel-silicon sensing layer, combined with sparse peripheral readout, transforms distributed optical dynamics into complementary temporal electrical signals at a zero-bias sensing front end. This architecture links self-driven photodetection, physical dimensionality reduction and temporal classification within a unified framework. By embedding part of the information processing directly into the sensing medium, polarization mediated vision offers a distinct route towards compact, task-oriented machine perception.

## Acknowledgement

### Funding

S. L. thanks the support from the National Natural Science Foundation of China (Nos. 51202216, 51551203, 61774135, and 62474161), the Distinguished Youth Fund of Zhejiang Natural Science Foundation of China (LR21F040001).

### Author contributions

S. L. conceived the study, designed and coordinated the overall experiments, participate in the experimental work, and wrote the paper. K. X. and M. Y. mainly carried out the experiments, wrote the paper and discussed the results. S. L. , M. Y., W. Z. and T. W. discussed the results and contributed to the experiments. X. B. and H. W. contributed to the development and implementation of the neural network. All authors contributed to the preparation of the manuscript.

**Competing interests**

The authors declare that they have no competing interests.